\documentclass[qsecnum,amsmath,preprintnumbers,superscriptaddress,nofootinbib,aps,prd,10pt,a4paper]{revtex4-1}
\pdfoutput=1

\usepackage[utf8]{inputenc}
\usepackage{lipsum}
\usepackage{amsmath}
\usepackage{amsfonts}
\usepackage{amssymb}
\usepackage{amsthm}
\usepackage[colorlinks=black, citecolor=green, linkcolor=blue, linktocpage=true]{hyperref}
\usepackage{xcolor}
\usepackage{bm}
\usepackage{graphicx}
\usepackage{comment}
\usepackage{mathrsfs}
\usepackage{caption}
\usepackage{subcaption}
\usepackage{framed}
\usepackage{microtype}   
\usepackage{array}
\def\baselinestretch{1.2}
\def\a{\alpha}
\def\b{\beta}

\def\m{\mu}
\def\n{\nu}

\begin{document}
\title{{\bf Superradiant scattering in Lorentz-violating gravity}}

\author{M. Herrero-Valea}
\email[]{mherrero@ifae.es}\thanks{Corresponding author.}

\address{Institut de Fisica d’Altes Energies (IFAE), The Barcelona Institute of Science and Technology, Campus UAB, 08193 Bellaterra (Barcelona) Spain}

\author{E. Simon-Felix}
\email[]{elesf01@gmail.com}

\address{Institut de Fisica d’Altes Energies (IFAE), The Barcelona Institute of Science and Technology, Campus UAB, 08193 Bellaterra (Barcelona) Spain}

\date{\today}

\begin{abstract}

Black holes in Lorentz violating gravity enjoy a double horizon structure which resembles that of the Kerr solution in General Relativity. Moreover, when a scalar field with a modified dispersion relation is coupled to these backgrounds, an on-shell mode with negative energy becomes possible in the region in between the horizons. Both properties together call for the possibility of extracting energy from the black hole by scattering of waves, even if the space-time is stationary. Here, we perform such scattering explicitly in the frequency domain, showing that indeed, a superradiant effect, leading to energy extraction, can be observed for modes with $l>0$ in spherical symmetry. In particular, we show that the mode $l=1$ can display a reflectivity exceeding 700\% for certain values of the Lorentz violating scale. This leads us to conjecture the instability of these space-times against such perturbations, although astrophysical size objects should have long life-times. Our results are not unique to Lorentz violating gravity and can be extended to any setting where modified dispersion relations are present, such as analogue gravity.

\end{abstract}

\maketitle
{\renewcommand{\baselinestretch}{1.2} \parskip=0pt
\setcounter{tocdepth}{2}
\tableofcontents}
\newpage

\section{Introduction}
The possibility of extracting energy from a black hole was first proposed by Penrose in the early 1970s \cite{Penrose:1971uk}, by noting that the Kerr solution exhibits two different types of horizons. Whereas for stationary black holes the event horizon, denoting the causal frontier of space-time, coincides with the Killing horizon -- the surface at which the Killing vector associated to energy conservation changes character --, this is not the case anymore when the black hole is rotating. In the latter, the Killing horizon always lays outside the inner event horizon, which thus allows for a narrow region of space-time, named ergosphere, where worldlines with negative Killing energy can be on-shell, and from which an observer can escape back to large radii. By leveraging these properties, it is kinematically possible for a body to enter the ergosphere and exit it with larger energy by leaving behind a piece of itself with negative energy, which falls into the event horizon of the black hole. As seen by an observer at spatial infinity, the body has stolen some of the rotational energy of the black hole.

A specific realization of this process is given by \emph{superradiance} \cite{Brito:2015oca}, a phenomenon where waves scattering against a geometry can exhibit reflection coefficients larger than one. This is the case, for instance, of scalar waves of light massive fields around Kerr-like solutions, which can manifest superradiance if their frequency $\hat \omega$ satisfies $\hat \omega< m O_H$, where $m$ is the mass of the field and $O_H$ is the angular velocity of the Killing horizon. This leads to the development of a field cloud around the black hole, and even to an explosive extraction of the rotational energy in some cases -- see \cite{Brito:2015oca} and references therein.

The key property that allows for superradiance in black hole geometries is the aforementioned possibility for on-shell trajectories to move in the region behind a Killing horizon and still escape back to its exterior. This scenario can be realized in some theories beyond General Relativity (GR) by allowing propagating modes to travel at speeds larger than the maximal speed set by the metric. This is the case, for instance, of some models of scalar-tensor gravity within the Hordensky and beyond Hordensky families \cite{Kobayashi:2019hrl}. Although most of these possibilities are ruled out by gravitational wave observations \cite{Ezquiaga:2017ekz}, there is a still a possible approach which is compatible with all observational results, consisting on abandoning boost invariance and thus Local Lorentz Invariance (LLI) as a defining symmetry \cite{Gupta:2021vdj}. This allows naturally for superluminal speeds and can be easily realized by coupling matter fields to a time-like unit vector, the aether, which imprints them with a modified momentum-dependent dispersion relation. The dynamics of the aether is then dictated by Einstein-Aether (EA) gravity \cite{Jacobson:2000xp, Jacobson:2004ts}, which incorporates LLI violations within a covariant framework.

Black holes in EA gravity have a double horizon structure similar to that of the Kerr solution in GR \cite{Berglund:2012bu, Cropp:2013sea}. The usual Killing horizon becomes permeable to fields that couple to the aether, that allows them to exceed the speed of light. However, there is an inner surface, named \emph{universal horizon}, that traps motions of any speed, due to the causal properties of the foliation orthogonal to the aether. In the intermediate region, the character of the Killing vector flips from being time-like in the exterior of the Killing horizon, to space-like. As a consequence, modes with negative energy can be created on-shell \cite{DelPorro:2022vqi}. This construction reproduces all the features needed to exploit a Penrose-like process.

The possibility of extracting energy from black holes through Lorentz violations has been previously studied in several works \cite{Eling:2007qd,Jacobson:2010fat,Cardoso:2024qie}, concluding that the possibility was as most marginal. However, all these works only consider fields travelling with quadratic dispersion relations of the form $\omega^2 = c^2k^2$ with different values of $c$. Although this can be realised in several ways, the presence of the aether in the setting at hand allows for many other possibilities. In particular, we focus here on the consequences of adding higher-derivative operators along spatial directions to the matter fields. This leads to modified dispersion relations $\omega^2 = f(k)$ with a polynomial function $f(k)$. By coupling these fields to EA gravity, we can embed the whole construction within a covariant theory and show that, for certain choice of the parameters of the theory, the scattering of matter waves against the geometry exhibits superradiance in a way which is close to the phenomenology of Kerr black holes.

This specific choice of operators might seem exotic and not well justified, but it is actually inspired by Ho\v rava gravity \cite{Horava:2009uw,Blas:2010hb}, a power-counting renormalizable theory of gravitation where LLI violations are a built-in feature \cite{Barvinsky:2015kil}. The action of Ho\v rava gravity can be rewritten as that of EA gravity -- with an extra condition of hypersurface orthogonality for the aether vector -- plus a set of higher derivative operators constructed only along the directions orthogonal to the aether. Any matter field coupled to such a theory is expected to develop the same kind of operators through radiative corrections \cite{Barvinsky:2017mal,LopezNacir:2011mt}. Therefore, here we assume implicitly that our UV completion to be Ho\v rava gravity, although our results can be extended beyond this specific case and apply also to other situations with modified dispersion relations and, in particular, to analogue models of gravitation.

This paper is organized as follows. First, we introduce EA gravity, together with stationary black hole solutions in section \ref{sec:black_holes}. Then, we move to discuss the dynamics of a scalar field equipped with higher-derivative LLI violating operators in section \ref{sec:Lifshitz_field}, arguing that superradiance should be possible within this setting in \ref{sec:superradiance}. Later, we introduce our strategy to compute a explicit scattering in section \ref{sec:scattering}, that we perform through the numerical strategy explained in \ref{sec:numerical}. Finally, we discuss our results in section \ref{sec:results} and draw conclusions in section \ref{sec:conclusions}. We also provide an appendix containing the explicit form of the coefficients appearing in the differential equations that we solve.

\section{Black holes in Einstein-Aether gravity}\label{sec:black_holes}
In order to embed LLI violations withn a covariant setting, ee focus on fields coupled to EA gravity \cite{Jacobson:2000xp}, with action\footnote{Hereinafter we use a mostly minus signature.}
\begin{align}\label{eq:EA_action}
    S_{\rm EA} = \frac{1}{16 \pi G}\int d^4x \sqrt{|g|}\left(-R+K_{\m\n}^{\a\b} \nabla_\a U^\m \nabla_\b U^\n + \zeta(U^\m U_\m -1)\right),
\end{align}
where $G$ is the Newton's constant, $R$ the Ricci scalar, $U^\m$ the aether vector, $\zeta$ is a Lagrange multiplier imposing the unit norm and time-like character of the aether, and the tensor $K_{\m\n}^{\a\b}$ is defined by
\begin{align}
    K_{\m\n}^{\a\b} = c_1 g^{\a\b} g_{\m\n} + c_2 \delta^\a_\m \delta^\b_\n + c_3 \delta^\a_\n \delta^\b_\m + c_4 U^\a U^\b g_{\m\n},
\end{align}
with the four $c_i$ being couplings that characterize the theory in parameter space. 

As it is written here, EA gravity is the most general theory of Lorentz violating gravity with up to two derivatives, and can thus be thought as a low energy effective field theory to describe Lorentz violating phenomena in gravitation. In particular, it contains the low energy limit of Ho\v rava gravity, also known as Khronometric Gravity \cite{Horava:2009uw,Blas:2010hb}. Indeed, the action of EA gravity matches that of Ho\v rava gravity when restricted to second derivatives as long as we enforce the Frobenius condition $U_{[\m}\nabla_\n U_{\rho]}=0$, which implies that the aether is hypersurface orthogonal, thus defining a co-dimension one foliation.

The action \eqref{eq:EA_action} admits asymptotically flat, spherical and static black hole solutions for generic values of the couplings $c_i$ through the ansatz \cite{Barausse:2011pu}
\begin{align}
    &ds^2 = F(r)dt^2 - \frac{B(r)^2}{F(r)}dr^2 - r^2 dS^2,\\
    &U_\m dx^\m = \frac{1+F(r)A(r)^2}{2A(r)} dt + \frac{B(r)}{2A(r)}\left(\frac{1}{F(r)}-A(r)^2\right)dr,
\end{align}
where $F(r)$, $A(r)$ and $B(r)$ are functions of the radial coordinate and $dS^2 = d\theta^2 + \sin^2 \theta\ d\phi^2$ is the usual $S^2$ volume element. The specific form of $U_\m$ here is chosen for computational convenience, and satisfies the unit norm constraint automatically. Coincidentally, the spherically symmetric condition on $U^\m$ automatically enforces hypersurface orthogonality, thus defining a preferred foliation in terms of space-like hypersurfaces orthogonal to it. Therefore, these space-times are also solutions of Khronometric gravity, which justifies our choice of Ho\v rava gravity as our UV completion.

Plugging this ansatz into the equations of motion derived from \eqref{eq:EA_action}, a solution for the radial functions can be found at any point of the parameter space \cite{Barausse:2011pu}. However, most of them are only accessible numerically. Only in two very specific cases we can obtain an analytical solution \cite{Berglund:2012bu}. Here we focus on the simplest of them, corresponding to $c_1+c_2+c_3 = 0$, where we add the extra condition $c_4=c_1+2 c_3$ in order to achieve an even simpler form of the metric. In this case
\begin{align}
    F(r) = 1-\frac{r_0}{r}, \quad B(r) = 1,\quad A(r) = 1.
\end{align}

As a consequence, the space-time solution that we study through this work corresponds to a Schwarzschild black-hole appended by the aether configuration, where $r_0$ is the size of the Killing horizon, identified by $\chi^2=F(r)=0$, with $\chi^\m$ the time-like Killing vector $\chi_\m dx^\m= F(r) dt$
\begin{align}
    &ds^2 = \left(1-\frac{r_0}{r}\right)dt^2 - \frac{1}{1-\frac{r_0}{r}}dr^2 - r^2 dS^2,\\
    &U_\m dx^\m = \left(1-\frac{r_0}{2r}\right) dt + \frac{r_0}{2r}\frac{dr}{1-\frac{r_0}{r}}.
\end{align}

We must remark here, however, that this specific solution has been ruled out by recent observational bounds \cite{Gupta:2021vdj}. Nevertheless, we expect the general properties of the solutions in other regions of the parameter space to be tantamount to those exhibited by this one, albeit implying much more challenging computations. Therefore, we stick here to the simplest analytic choice as a proof of concept for the results discussed below.

Due to the presence of the foliation, causality in this space-time is given in terms of the flow of time defined by $U^\m$. It is then natural to align coordinates with this structure, by performing a change of variables to the proper time of the foliation
\begin{align}\label{eq:change_var}
    d\tau = dt +\frac{dr}{F(r)}\frac{1-F(r)}{1+F(r)},
\end{align}
so that the metric and aether now read
\begin{align}\label{eq:space-time}
    &ds^2 = \left(1-\frac{r_0}{r}\right)d\tau^2 -\frac{r_0}{r}\frac{dtdr}{1-\frac{r_0}{2r}} -\frac{dr^2}{\left(1-\frac{r_0}{2r}\right)^2}dr^2 - r^2 dS^2,\\
    & U_\m dx^\m = \left(1-\frac{r_0}{2r}\right)d\tau,
\end{align}
while the Killing vector has become
\begin{align}
    \chi_\m dx^\m = \left(1-\frac{r_0}{r}\right)d\tau -\frac{r_0}{2r}\frac{dr}{1-\frac{r_0}{2r}}.
\end{align}
Notice that the new metric is independent of $\tau$ and thus evolution along preferred time will also conserve the Killing energy associated to $\chi^\m$.

After this transformation, the metric takes the Arnowitt-Deser-Misner (ADM) form
\begin{align}
    ds^2 = (N^2-N_i N^i)d\tau^2 -2N_i dx^i d\tau - h_{ij}dx^i dx^j,
\end{align}
where $N$, $N^i$, and $h_{ij}$ are the lapse, shift, and induced metric in the foliation leafs, respectively, given by
\begin{align}
    N = 1-\frac{r_0}{2r},\quad N_i dx^i = \frac{r_0}{2r}\frac{1}{1-\frac{r_0}{2r}}, \quad h_{ij}dx^i dx^j = \frac{dr^2}{\left(1-\frac{r_0}{2r}\right)^2} + r^2 d\Omega^2.
\end{align}

\begin{figure}
    \centering
    \includegraphics[width=0.7\linewidth]{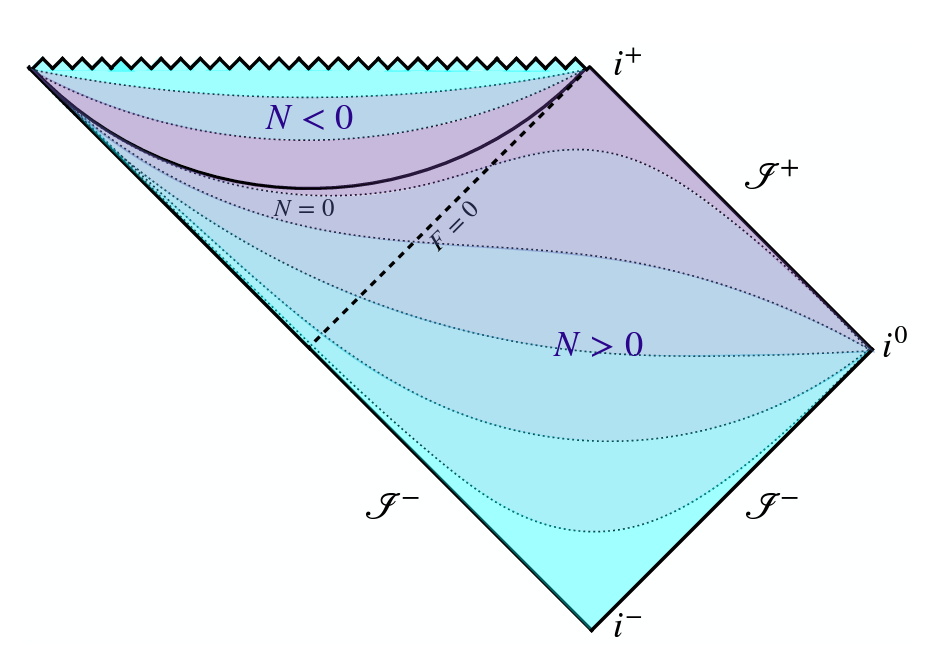}
    \caption{Penrose diagram of the geometry. The dotted lines correspond to the foliation leafs, while the color code -- from blue to magenta -- indicates the value of the proper time. The Killing horizon is shown as a dashed line, the UH as a solid line at the position $N=0$, and the singularity as a zigzag line. The asymptotic regions are defined in the same way as in plain GR. Figure from \cite{DelPorro:2023lbv}.}
    \label{fig:Penrose}
\end{figure}

Notice also that this chart of coordinates has an important pathology whenever $N=0$, which corresponds to $r=r_0/2$. From \eqref{eq:change_var}, we see that this lays at finite $r$ but corresponds to $\tau \rightarrow +\infty$, signalling that the foliation cannot be extended beyond this point. However, this surface can be crossed smoothly, since the proper time of an observer moving with the foliation, given by $ds_{\rm proper}=U_\m dx^\m$, remains finite. Hence, the surface $r_{\rm UH}=r_0/2$ signals a trapped region, beyond which no motion, regardless of their velocity, can escape, since the product $(\chi\cdot U)$ flips sign precisely at this point. For this reason, this surface is named \emph{universal horizon} (UH). In contrast, the Killing horizon at $r=r_0$ loses its meaning as a trapping surface. Any trajectory which is coupled to the aether will be able to cross it and get back to its exterior without any issue \cite{DelPorro:2023lbv}. However, its time(space)-like character is preserved, and its associated Killing vector still defines a conserved quantity, which will be relevant in what follows. For a more detailed discussion on the causal aspects of these space-times, see \cite{Bhattacharyya:2015gwa, DelPorro:2022kkh}. A Penrose diagram for the case at hand is shown in figure \ref{fig:Penrose}. 

Finally, as it will be useful later, let us also introduce a change of variables that implements the gauge choice $N^i=0$ while maintaining the ADM structure of the metric, by defining
\begin{align}
    d\rho = dt + \frac{4 r^2}{r_0}\frac{dr}{2r-r_0},
\end{align}
whose generator corresponds to 
\begin{align}
    S_\m dx^\m = -\frac{r_0}{2r}dt - \frac{2r}{2r-r_0}dr,
\end{align}
and spans a `radial' direction orthogonal to the aether time direction, since $S^2 = -1$ and $(S\cdot U)=0$.

\section{The covariant Lifshitz field}\label{sec:Lifshitz_field}
We will couple to this geometry a scalar field of the Lifshitz kind \cite{Ardonne:2003wa, Rubio:2023eva}, with action
\begin{align}\label{eq:action_scalar}
    S_{\phi}= \frac{1}{2}\int d^4x\sqrt{|g|}\left(\partial_\tau \phi \partial_\tau \phi -\phi (-\Delta_\gamma)\phi-\frac{\lambda}{\Lambda^2}\phi (-\Delta_\gamma)^2\phi\right),
\end{align}
where $\Delta_\gamma=\gamma^{\m\n}\nabla_\m \nabla_\n$ is the Laplatian operator along the directions orthogonal to $U^\m$, $\gamma_{\m\n}=-g_{\m\n}+U_\m U_\n$ is the orthogonal projector, and $\partial_\tau \equiv U^\m\nabla_\m \phi$ is the derivative along $U^\m$, and hence along the preferred time direction set by the foliation. Note that the first two terms in this action combine into the usual D'Alembert operator and thus correspond to the relativistic invariant action of a scalar field in $3+1$ dimensions. The last operator, instead, breaks LLI by coupling the field to the aether. Here $\lambda$ is a dimensionless coupling and $\Lambda$ denotes the energy scale at which the dynamics introduced by Lorentz violations become important. Note that in principle we could have introduced a whole tower of irrelevant LLI violating operators by using higher powers of $(-\Delta_\gamma)$, but we decided to retain only the lowest of these operators, in an effective field theory way of thinking, and also in order to simplify our computations. 

Let us also notice that although in an arbitrary chart of coordinates this action will contain four time derivatives, thus leading to Orstrogradsky ghosts, this is not the case when time evolution is taken along the aether direction \cite{Blas:2009yd}. In preferred frame coordinates, $(-\Delta_\gamma)$ is purely spatial and leads only to a momentum dependence of the dispersion relation -- cf. later -- but not to ghost modes.

Last, but not least, let us highlight that this particular family of UV operators is motivated by the construction of Ho\v rava gravity, that we take here implicitly as our UV completion, and whose Lagrangian displays terms constructed with $(-\Delta_\gamma)$ acting on the metric degreees of freedom. Whenever a scalar field is coupled to it, we expect radiative corrections to generate equivalent operators \cite{Barvinsky:2017mal, LopezNacir:2011mt}, which justifies our choice.


\subsection{Superradiance?}\label{sec:superradiance}
The dual horizon structure of the space-time \eqref{eq:space-time} resembles that of the Kerr solution in GR. We have a permeable Killing horizon, and an inner event horizon, which in this case is the UH. Moreover, rays of the field $\phi$ can move at speeds larger than the speed of light, due to the modified dispersion relation implied by \eqref{eq:action_scalar}. This can be seen immediately by focusing on spherically symmetric configurations of the field $\phi(\tau,r)$ and using a WKB ansatz for the solution
\begin{align}
    \phi \propto e^{i {\cal S}}, \quad {\cal S} =\int p_\m dx^\m,
\end{align}
where $p_\m$ is the momentum of the field $p_\m \phi = -i \nabla_\m \phi$. 

In terms of the coordinates of the preferred foliation $(\tau,\rho)$, we then introduce the preferred energy and momentum as $\omega = U^\mu p_\mu$ and $k = -S^\mu p_\mu$, so that
\begin{align}
    {\cal S}= \int \omega d\tau -\int k d\rho.
\end{align}

In the WKB limit $k\gg \nabla k$ this then leads to the following dispersion relation when plugged into the equations of motion of the scalar field
\begin{align}\label{eq:dispersion_relation}
    \omega^2 = k^2 + \frac{\lambda}{\Lambda^2}k^4,
\end{align}
which indeed induces a momentum dependent group velocity $v_g = (1+2 k^2 \lambda/\Lambda^2)/v_p$, where $v_p$ is the phase velocity\footnote{Note that in the decoupling limit $\lambda/\Lambda^2\rightarrow0$ we get $v_g=v_p^{-1}$, but also $\omega = k$, so that $v_g=v_p =1$. However, beyond leading order, the two velocities differ. For instance, the first correction in $\lambda/\Lambda^2$ leads to $v_p = 1+\lambda k^2/(2 \Lambda^2)$ and $v_g = v_p + \lambda k^2 /\Lambda^2$.} $v_p= \omega/k$.

\begin{figure}
    \centering
    \includegraphics[width=0.8\linewidth]{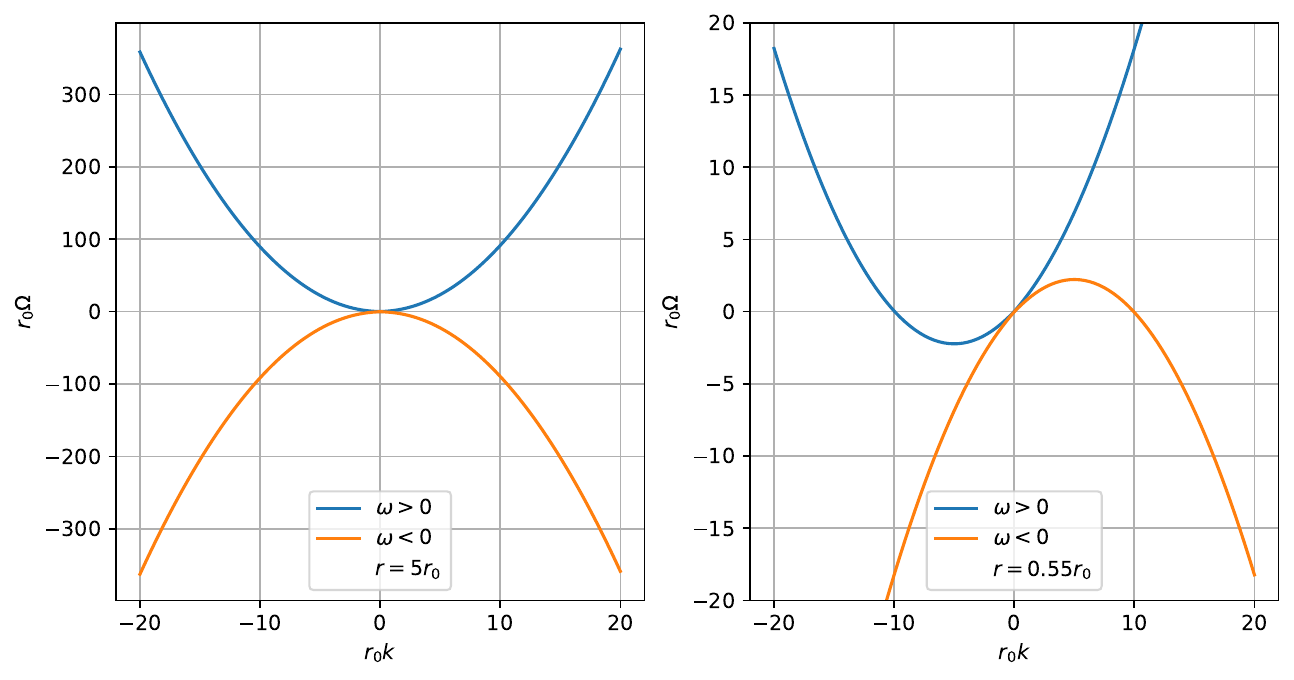}
    \caption{Dispersion relation at different points of the space-time for $\lambda/\Lambda^2 = r_0^2$. In the region between the two horizons, the parabolas are shifted and modes with negative energy can become on-shell.}
    \label{fig:dispersion}
\end{figure}

Note however that neither $U^\m$ nor $S^\m$ are Killing vectors. Thus, their associated momentum contributions are not conserved quantities and therefore the solutions of the dispersion relation \eqref{eq:dispersion_relation} will differ in different regions of the space-time. In order to make this explicit, it is convenient to introduce the conserved Killing energy $\Omega$, defined by $\Omega = k_\m \chi^\m$ so that we can write
\begin{align}\label{eq:Omega_relation}
    \Omega = \omega (U\cdot \chi)-k (\chi\cdot S), 
\end{align}
after decomposing the momentum vector along the preferred directions as $p_\m =\omega U_\m - k S_\m$. Together with \eqref{eq:dispersion_relation}, we now have two equations for two variables than can be solved at every point of space-time in terms of the contractions of the Killing vector $\chi^\mu$ with the generators of the preferred directions. This allows us to evaluate the physical character of the propagating modes, by examining the dispersion relation outside and inside the Killing horizon. As we can see in figure \ref{fig:dispersion}, while the dispersion relation looks similar to a relativistic one whenever $r > r_0$, containing only two solutions for positive Killing energy $\Omega$, this is not the case anymore when $r<r_0$. In the latter case, the parabolas shift their position, and negative energy modes can be produced on-shell.

\begin{figure}
    \centering
    \includegraphics[width=0.5\linewidth]{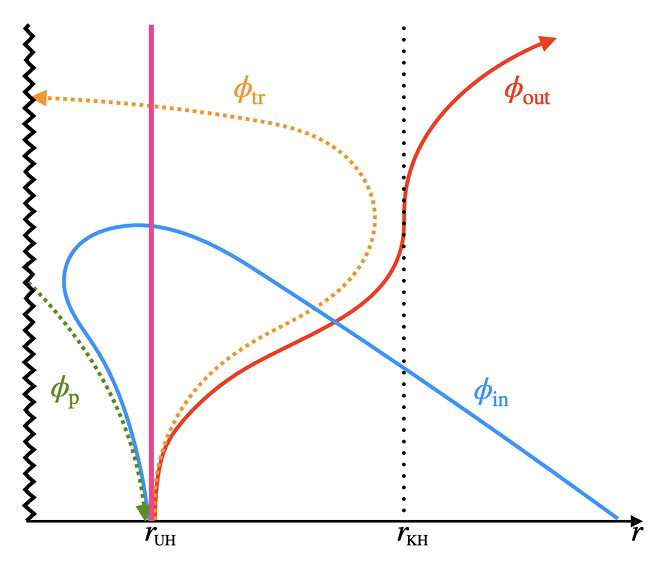}
    \caption{Cartoon picture of the characteristics describing the trajectories of the possible on-shell modes. Proper time grows in the positive vertical direction. The UH is represented by the magenta line, while the dotted line corresponds to the Killing horizon, and the zigzag line to the singularity. Adapted from \cite{DelPorro:2022vqi}}
    \label{fig:characteristics}
\end{figure}

A more detailed picture can be acquired by using the method of characteristics as in \cite{DelPorro:2022vqi, Michel:2015rsa}, which leads to the schematic structure shown in figure \ref{fig:characteristics}. The modes in the exterior of the black hole can be identified with the usual in-going and out-going modes also found in GR, which can be continued down the geometry. The in-going mode $\phi_{\rm in}$ crosses the UH smoothly, while the out-going mode $\phi_{\rm out}$ can be traced back to peel from the UH, instead of the Killing one, due to the presence of the modified dispersion relation. Still, for large values of $\Lambda$ -- corresponding to the decoupling limit --, this mode spends a long time at the Killing horizon, mimicking the general relativistic behavior. In the interior region, in between the horizons, we find the new two modes. One of the them crosses the UH smoothly, while the other peels from it. At some intermediate point, before reaching the Killing horizon, these two solutions degenerate into one and become complex. This not only signals that they cannot exist in the exterior region, but also that they correspond to two tails of the same physical trajectory $\phi_{\rm tr}$ which always remains trapped in the interior region. 

Note also that in the inside of the UH, the roles of the $\phi_{\rm in}$ and $\phi_{\rm tr}$ modes gets swapped, while a new $\phi_{\rm p}$ trajectory appears, coming from the singularity to the UH. This is due to the fact that these geometries are not globally oriented, because the foliation cannot be extended beyond the UH. In its interior we find instead a copy of the foliation with reversed orientation, indicating that the time flow is also reversed \cite{DelPorro:2022kkh}. This property is interesting for studying quantum effects \cite{DelPorro:2023lbv}, but is far from the scope of this work. From now on we will focus on the exterior of the UH only, forgetting about the avatars of its interior.

This structure points to the possibility of activating a Penrose-like process that can lead to superradiant scattering of waves of the scalar field. In the exterior, a ray with positive Killing energy $\Omega_1$ can be produced, corresponding to a ray of positive local energy $\omega_1$. Although $\omega$ is not constant, its sign is, since it controls causal propagation in preferred time \cite{Blas:2009yd}. This ray can then enter into the geometry and split into a escaping ray with $\omega_2>0$ and $\Omega_2>0$, \emph{and a trapped ray with} $\omega_3>0$ \emph{but} $\Omega_3<0$, thanks to the deformation of the dispersion relation. When the escaping ray arrives back to larger regions, it will have done it with a Killing energy $\Omega_2 = \Omega_1-\Omega_3 > \Omega_1$. Thus, a superradiant channel seems plausible through this phenomenon. Note however that it is only possible to produce the trapping mode $\phi_{\rm trap}$ up to a cut-off in the absolute value of its Killing energy, corresponding to the minimum of the $\omega>0$ parabola in fig \ref{fig:dispersion}. This seems to be the condition equivalent to $\hat \omega<m O_H$ found in the standard case of the Kerr solution in GR, and hints that superradiance should be possible only for those values of the Killing energy which sit below the cut-off.


\section{Scattering in the frequency domain}\label{sec:scattering}

After discussing the possibility of extracting energy from the black hole configuration in the case at hand, we perform now a explicit numerical computation of the scattering of scalar waves against the geometry. We work in the chart of coordinates given by $(\tau, r, \theta, \phi)$ and assume a decomposition of the scalar field in spherical harmonics of the form
\begin{align}
\label{ansatz}
    \phi(\tau, r, \theta, \phi) = \sum_{l,m}\int \frac{d\Omega}{2\pi}\ e^{i\Omega \tau} Y_{lm}(\theta, \phi) \frac{\psi_{lm}(r)}{r},
\end{align}
where $Y_{lm}(\theta, \phi)$ is the $l,m$ spherical harmonic, and we have also performed a Fourier transform in preferred time, thus working in frequency domain from now on.

Plugging this expression into the equations of motion derived from \eqref{eq:action_scalar}, we arrive at a single differential equation for $\psi(r)$ in the form
\begin{align}\label{eq:4_order_eq}
    \alpha_4(r,l)\frac{d^4 \psi_{lm}}{dr^4}+\alpha_3(r,l)\frac{d^3 \psi_{lm}}{dr^3}+\alpha_2(r,l)\frac{d^2 \psi_{lm}}{dr^2}+\alpha_1(r,l)\frac{d \psi_{lm}}{dr}+\alpha_0(r,l) \psi_{lm}(r)=0,
\end{align}
where the functions $\alpha_i(r,l)$ are given in the appendix, but importantly, they are independent of $m$ due to angular momentum conservation.

At large radii, this equation exhibits four solutions -- two propagating ones, corresponding to the in-going and out-going modes described previously, and two real exponentials. A generic solution when $r \rightarrow \infty$ is then
\begin{align}\label{eq:soft_solution}
    \psi(r) = I e^{i k_- r}+ R e^{-i k_- r}+ C_+ e^{k_+ r} + C_- e^{- k_+ r},  
\end{align}
with $I$ and $R$ the incidence and reflection coefficients, and
\begin{align}
    k_{\pm}= \left[\frac{\sqrt{1+4\epsilon \Omega^2}\pm 1}{2\epsilon}\right]^{\frac{1}{2}},
\end{align}
where we have defined 
\begin{align}
    \epsilon\equiv \frac{\lambda}{\Lambda^2}.
\end{align}
Note that all coefficients accompanying the different particular solutions are complex in general, and that the solution is independent of $l$ and $m$.

We are performing a scattering experiment, so we will fix the value of $I$ and measure its corresponding reflection coefficient $R$, whose value will indicate whether we find superradiance or not. The other two coefficients accompanying the real exponentials will be set to vanish, since they do not correspond to physical propagating modes. Altogether this provides three of the four boundary conditions required to solve \eqref{eq:4_order_eq}. The remaining condition, as we will see in a moment, will be set at the UH, which represents the inner causal boundary of the space-time. In practice, we will exploit the linearity of the equation to divide the solution by $I$, effectively choosing $I=1$ as our boundary condition. The coefficient of the negative imaginary exponential then becomes $\hat R = R/I$ and will directly provide the quantity that we aim to compute.

In the neighborhood of the UH we also find four solutions, which correspond to the modes displayed in figure \ref{fig:characteristics}. Their behavior when approaching the UH at $r_{\rm UH} =r_0/2$ differs and can be studied using boundary layer theory. We change variables to $ k= \delta \hat k$ and look for values of $\delta$ that compensate different terms of the dispersion relation, solving it for all of them at leading order in $\epsilon$. We find only two non-trivial cases
\begin{itemize}
    \item Soft modes $\delta = \epsilon^{-1/2}$ :
    \begin{align}
        \hat k = -\frac{\Omega}{(S\cdot \chi)}, \quad \omega = \sqrt{k^2 + \epsilon k^4}.
    \end{align}
    These are modes that cross the UH smoothly, with their local momentum and energy remaining finite at that position. They thus correspond to $\phi_{\rm in}$ and to the soft tail of the trapped mode.

    \item Hard modes $\delta = \frac{(S\cdot \chi)}{\sqrt{\epsilon} (U\cdot\chi)}$ :
    \begin{align}\label{eq:hard_modes_wk}
        \hat k = \pm \delta, \quad \omega = \pm \frac{\delta}{\sqrt{\epsilon}}.
    \end{align}
    Both the momentum and local energy of these modes diverge when approaching the UH, since $(U\cdot \chi)$ vanishes there. They thus correspond to $\phi_{\rm out}$ -- for the negative sign, since $\delta<0$ -- and the hard tail of the trapped mode -- for the positive sign.
\end{itemize}

The next step is to evaluate these explicitly in the neighborhood of the UH. We start with the soft modes, by proposing a simple ansatz of the form 
\begin{align}
    \psi_{\rm soft}(r) = \left(r-\frac{r_0}{2}\right)^b +{\cal O}\left[\left(r-\frac{r_0}{2}\right)^{c} \right] ,
\end{align}
with ${\rm Re}(c)>{\rm Re}(b)$, in order to determine the leading behavior of the functions. Plugging this into \eqref{eq:4_order_eq} and expanding to leading order in $(r-r_0/2)$ we obtain $b_1= ir_0 \Omega/2$, and $b_2 = b_1 + 1$.

The hard modes, as we can see in \eqref{eq:hard_modes_wk}, have energies and momenta which are singular when $\epsilon \rightarrow 0$. Hence, in order to obtain them we need to resort to singular perturbation theory, proposing
\begin{align}
    \psi_{\rm hard}(r) = \exp\left(\frac{y(r)}{\sqrt{\epsilon}}\right) + {\cal O}\left(\sqrt{\epsilon}\right),
\end{align}
and assume a power series form for $y(r)$ similar to that of the soft modes. Again, plugging this into \eqref{eq:4_order_eq} we find two solutions at leading order -- besides the trivial one $y(r)=0$
\begin{align}
    y_{\pm}(r)=\pm\frac{i r_0^2}{4r-2r_0}+{\cal O}\left[(r-\frac{r_0}{2})^{0} \right],
\end{align}
where the plus (minus) sign corresponds to the out-going (trapped) mode, as can be seen by evaluating $\omega$ for each of them. 

Hence, we conclude that the solution in the neighborhood of the UH can be written as
\begin{align}\label{eq:UH_solutions}
    \psi(r) = B_1 \psi_{{\rm soft}1}+B_2 \psi_{{\rm soft}2}+H_+ \psi_{{\rm hard}+}(r)+H_- \psi_{{\rm hard}-}(r),
\end{align}
which is again independent of $l$ and $m$.

We are now finally able to come back to the issue of fixing boundary conditions for our scattering problem. We had already three conditions defined at large radii, thus we can only fix another one. Since the UH is a semi-permeable trapping surface, we do not want any emission from it to reach infinity, so the obvious choice is $H_+ = 0$, which kills the out-going mode. Although the trapped mode is also generated from the UH, it never escapes to infinity, and does not modifies the energy balance of the horizon, as it always falls back into it. Therefore, there is no reason as to why it should not exist. We finally have our four boundary conditions for the scattering problem
\begin{align}\label{eq:boundary_conds}
    I=1,\quad C_+=C_-=H_+=0.
\end{align}

\begin{figure}
    \centering
    \includegraphics[width=0.5\linewidth]{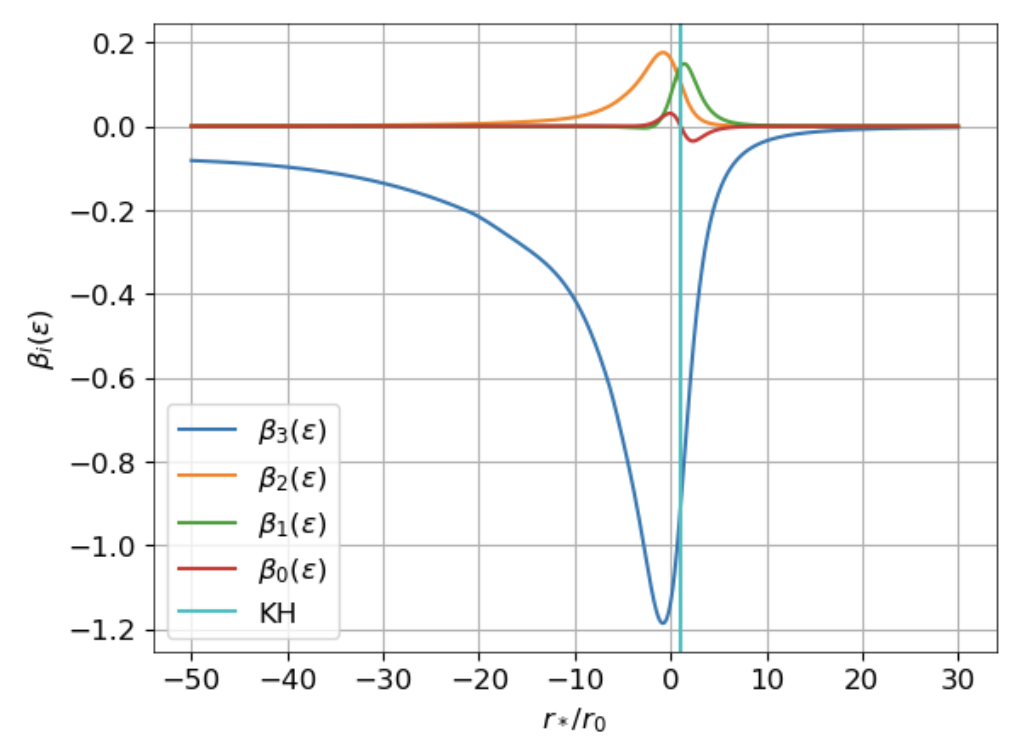}
    \caption{Terms proportional to $\epsilon$ in the coefficients $\beta_i(r,l)$ for $r_0=1, \Omega=1.5, \epsilon=0.1$ and $l=0$. We denote them $\beta_i(\epsilon)$.}
    \label{fig:higher-terms}
\end{figure}

\subsection{Numerical implementation}\label{sec:numerical}
 The next step is to obtain the behaviour of the scalar waves scattered against the geometry. We will do so by resorting to numerical techniques. Therefore, a first step is to decompactify our radial direction, by placing the UH at an infinite distance. In order to do this, we simply introduce a sibling of the standard tortoise coordinate\footnote{Note that any power of $(r-r_0/2)$ would serve the purpose here. We choose a quadratic power in order for the derivatives of the hard modes, which need to be evaluated at the UH in our algorithm, to remain ${\cal O}(1)$.}
 \begin{align}
       \frac{dr}{dr_*}=\left(1- \frac{r_0}{2r}\right)^2,
    \label{eq:tortoise_change}
\end{align}
which leads to
\begin{align}
     r_*=\frac{2r(r-r_0)}{2r-r_0} + r_0\log(2r-r_0),
    \label{eq:tortoise_coord}
\end{align}
so that the UH corresponds to $r_*\rightarrow -\infty$. Note that, as with the usual general relativistic tortoise coordinate, at large radii we have $r_* \sim r$.

\begin{figure}
    \centering
    \begin{subfigure}[b]{0.45\textwidth}
        \centering
        \includegraphics[width=\textwidth]{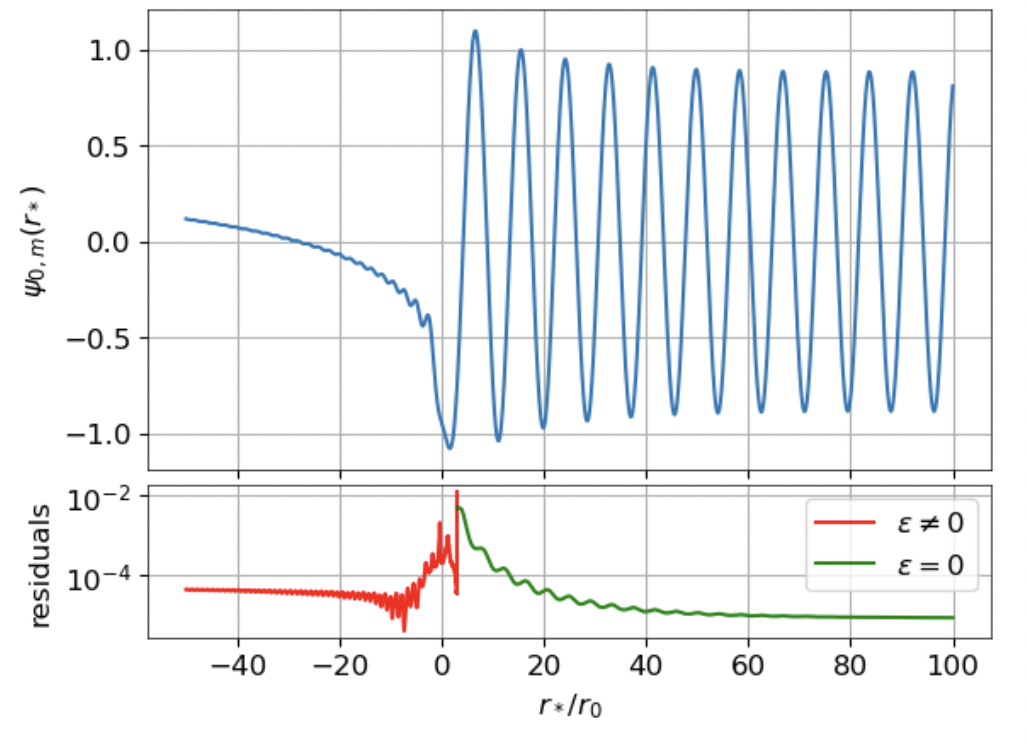}
        \caption{Benchmark solution for $l=0, \epsilon = 10^{-1} r_0^2$ and $r_0 \Omega= 0.75$. The inferior panel displays the residual value of the equation of motion at each point.}
        \label{fig:solutions_l_0}
    \end{subfigure}
    \hfill
    \begin{subfigure}[b]{0.45\textwidth}
        \centering
        \includegraphics[width=\textwidth]{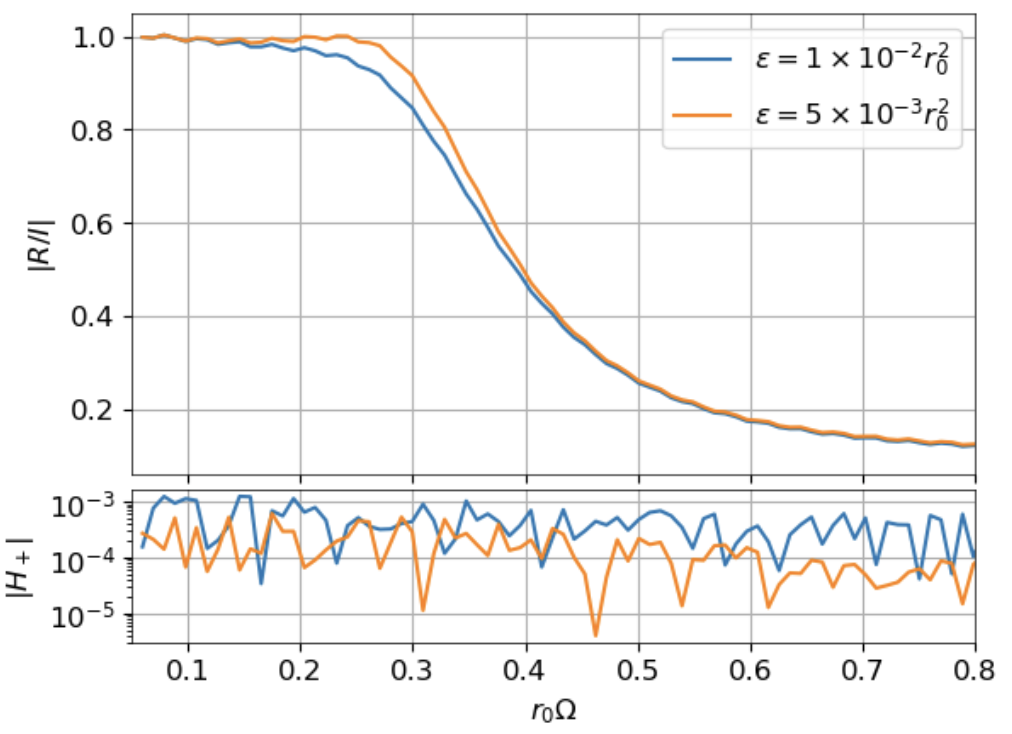}
        \caption{Ratio $|R/I|$ for different values of the Killing energy $\Omega$. The inferior panel displays the value of $|H_+|$ for the solution reached at each value of $\Omega$.}
        \label{fig:scan_l_0}
    \end{subfigure}
    \caption{Dynamics of the fundamental mode $l=0$. The left plot shows a benchmark solution, while the right plot displays the curve for $|R/I|$ obtained from our experiments.}
    \label{fig:l0}
\end{figure}

In terms of this new chart of coordinates, we thus rewrite \eqref{eq:4_order_eq} as
\begin{align}
      \label{eq:eq_tortoise}
    \frac{d^4 \psi_{lm}}{dr_*^4}+\beta_3(r,l)\frac{d^3 \psi_{lm}}{dr_*^3}+\beta_2(r,l)\frac{d^2 \psi_{lm}}{dr_*^2}+\beta_1(r,l)\frac{d \psi_{lm}}{dr_*}+\beta_0(r,l) \psi_{lm}(r_*)=0,  
\end{align}
where the coefficients $\beta(r,l)$ are given implicitly in terms of $r(r_*)$, and their explicit form can be found in the appendix.

We proceed by shooting in the complex coefficient $\hat R$. We integrate from large radii using a fourth order Runge-Kutta method, starting from the boundary conditions \eqref{eq:boundary_conds} and a randomly chosen value for $\hat R$. We will propagate the solution down to the UH and project it in the basis of solutions \eqref{eq:UH_solutions}, extracting the value of the coefficient $H_+$. In general, this will not vanish, signaling a failing of the fourth boundary condition in \eqref{eq:boundary_conds}. However, this procedure defines an implicit function $H_+(\hat R)$ whose root leads to the desired value. We will look for this root by using gradient descent, by minimizing the following function
\begin{align}
    \chi(\hat R) = |H_+(\hat R)|^2,
\end{align}
whose global minimum agrees with the solution of $H_+(\hat R)=0$.

In practice, however, the presence of higher derivatives in \eqref{eq:eq_tortoise} threatens the global stability of the numerical scheme. At large radii, the function $e^{-k_k r}$ in \eqref{eq:soft_solution} vanishes effectively and thus the integration scheme cannot distinguish between a solution with $C_-=0$ and another one where the coefficient does not vanish at this point. As a consequence, when starting our integration from large values of $r$, the real exponential mode gets triggered and dominates the solution, leading to a quick divergence as we move inwards in the radial direction. To achieve a successful integration, one must then ensure that this mode is absent \emph{at all values of $r$}, which is not an easy task.

In order to solve this issue, we proceed here by noting that the terms proportional to $\epsilon$ in \eqref{eq:eq_tortoise}, thus coming from the higher derivative operator in \eqref{eq:action_scalar}, only have a sizeable value once we approach the position of the Killing horizon, as it can be seen in figure \ref{fig:higher-terms}. Therefore, we will drop these terms for values of $r>3 r_0$, turning them on only after reaching this position. In practice this means that we will start by solving the second order general relativistic equation obtained from setting $\epsilon =0$ -- albeit using the tortoise coordinate \eqref{eq:tortoise_coord} -- from large radii, corresponding to $r_*/r_0=100$, and starting from a boundary condition
\begin{align}
    \psi(r) = e^{i \Omega r}+ \hat R e^{-i \Omega r}. 
\end{align}

Once the integrator has reached the value $r=3 r_0$, we then stop and use the solution at this point as initial condition to integrate now the full equation \eqref{eq:eq_tortoise} down to the UH, which in practical terms means $r_*/r_0 \sim -50$. Note that at all times this implies that $\epsilon$ must satisfy $\Omega^2 \epsilon\ll 1$ in order for the boundary condition to be consistent with the limit $\epsilon \rightarrow 0$.

\begin{figure}
    \centering
    \begin{subfigure}[b]{0.45\textwidth}
        \centering
        \includegraphics[width=\textwidth]{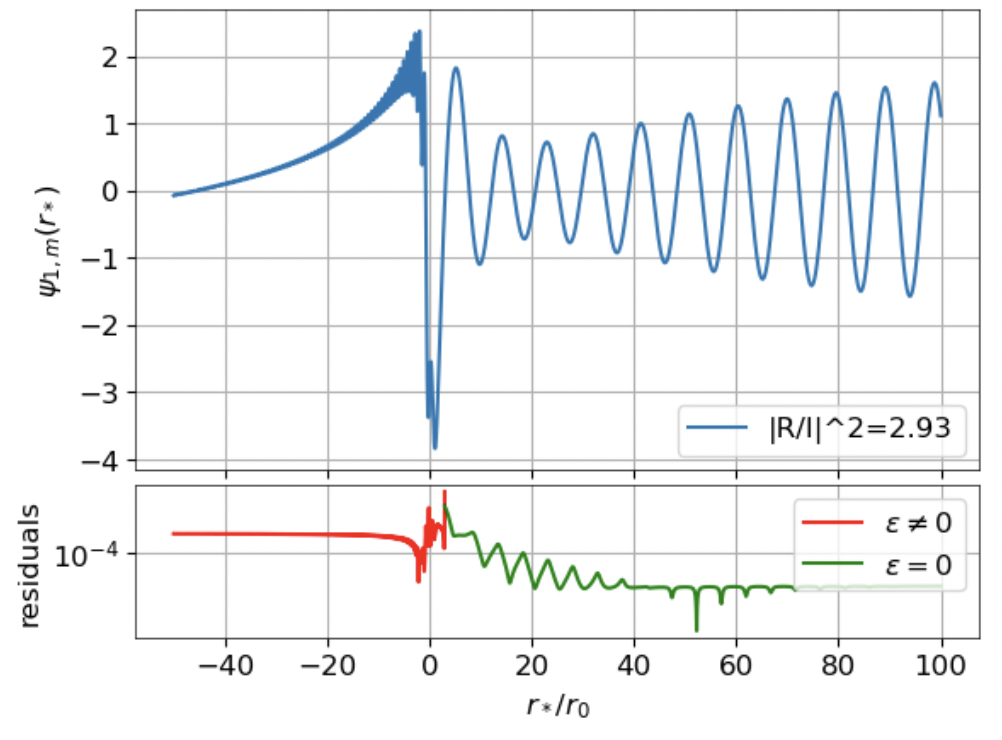}
        \caption{Solution displaying superradiance, obtained with $\epsilon = 10^{-2} r_0^2$ and $r_0 \Omega= 0.65$.}
        \label{fig:solution_SR}
    \end{subfigure}
    \hfill
    \begin{subfigure}[b]{0.45\textwidth}
        \centering
        \includegraphics[width=\textwidth]{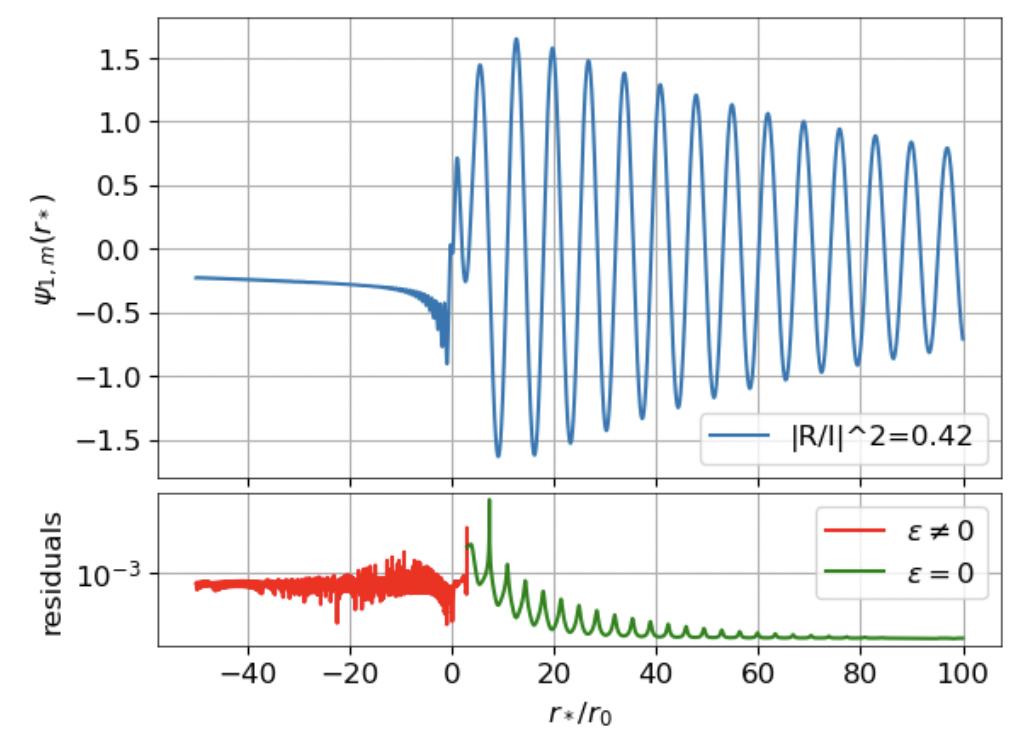}
        \caption{Solution without superradiance, obtained with $\epsilon = 10^{-2} r_0^2$ and $r_0 \Omega= 0.9$}
        \label{fig:solution_no_SR}
    \end{subfigure}
    \caption{Two example solutions, with and without superradiance, for $l=1$. Again, the change in the behavior of the curve signals the Killing horizon. The inferior plots display the value of the residuals of the equation of motion.}
    \label{fig:solution_l_1}
\end{figure}

\section{Results and discussion}\label{sec:results}

Following the algorithm described in the previous section, we are able to find solutions to \eqref{eq:eq_tortoise} with large accuracy and fourth order convergence order equivalent to that of the Runge-Kutta algorithm, except for the matching region $r\sim 3r_0$, where convergence is controlled instead by the finite differences formula used to compute the derivatives of the solution at the matching region. We perform all our experiments by fixing $r_0$ and varying the values of $\epsilon$, which controls the strength of the Lorentz violating terms in the equations, and of the Killing energy $\Omega$.

We first studied the simplest case of spherical waves, corresponding to $l=0$. A benchmark solution is shown in figure \ref{fig:solutions_l_0}. However, in this case we find no superradiance at all, as it can be confirmed by scanning the parameter space, as shown in figure \ref{fig:scan_l_0}. The situation is different for higher modes. In figure \ref{fig:solution_l_1} we display two solutions, with and without superradiance, for the case $l=1$. We observe that in the superradiant case, the wave in the interior of the Killing horizon is substantially enhanced with respect to the exterior wave. This is not the case when $|R/I|<1$, signaling that in the former, the modes in the interior are excited in order to provide the superradiant effect.

\begin{figure}
    \centering
    \begin{subfigure}[b]{0.45\textwidth}
        \centering
        \includegraphics[width=\textwidth]{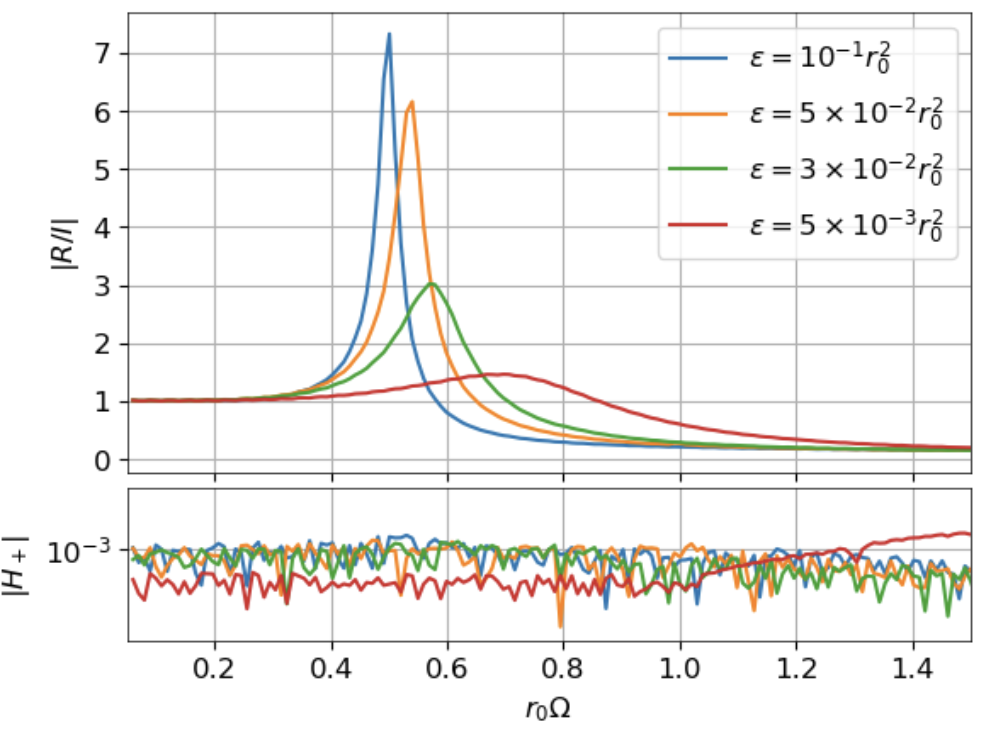}
        \caption{Result for $l=1$. We observe that a larger superradiant peak develops for increasing values of $\epsilon$.}
        \label{fig:scan_l1}
    \end{subfigure}
    \hfill
    \begin{subfigure}[b]{0.45\textwidth}
        \centering
        \includegraphics[width=\textwidth]{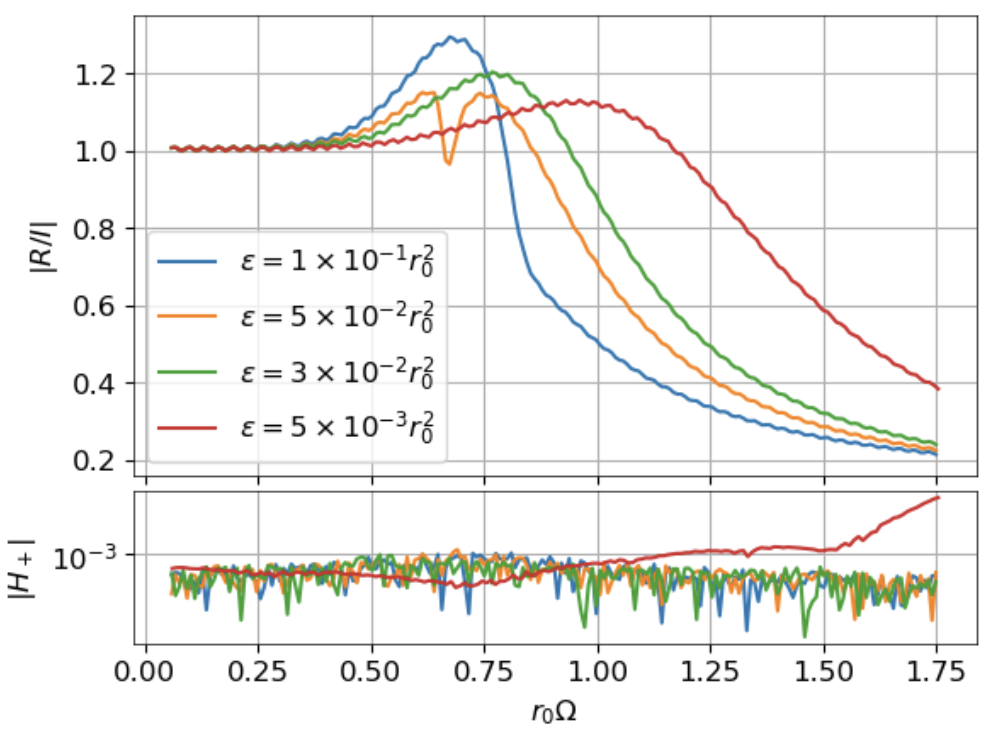}
        \caption{Results for $l=2$. Here we also find superradiance, but the values of $|R/I|$ are much more moderate.}
        \label{fig:scan_l2}
    \end{subfigure}
    \caption{Results for the coefficient $|R/I|$ in the cases $l=1$ and $l=2$. The inferior plot shows the residual value of $|H_+|$ for each Killing energy.}
    \label{fig:scan_l1_l2}
\end{figure}

Performing a scan of the parameter space for $l=1,2$, we observe that while for small values of $\epsilon \lesssim 10^{-2} r_0^2$, the curve for $|R/I|$ is similar to that encountered in a general relativistic setting for the Kerr solution, the situation becomes more drastic as the value of $\epsilon$ increases, and the Lorentz violation scale is reduced. As we can see, for $\epsilon \sim 10^{-1} r_0$ we observe a peak of $|R/I|\sim 7$ in the $l=1$ mode. This seems in tune with the results of \cite{Rubio:2023eva, Oshita:2021onq}, where the time evolution of an equivalent system is done, observing also a stronger cascade of scalar field outside the KH when the value of $\epsilon$ is increased. Notice that the curve always start at $|R/I|=1$ for $\Omega \rightarrow 0$, complying with the fact that large wavelengths completely miss the black hole and return to the boundary totally reflected -- keep in mind that we are working in spherical symmetry --, while for large $\Omega$ it vanishes, indicating that small wavepackets get completely absorbed by the black hole.

\begin{figure}
    \centering
    \includegraphics[width=0.5\linewidth]{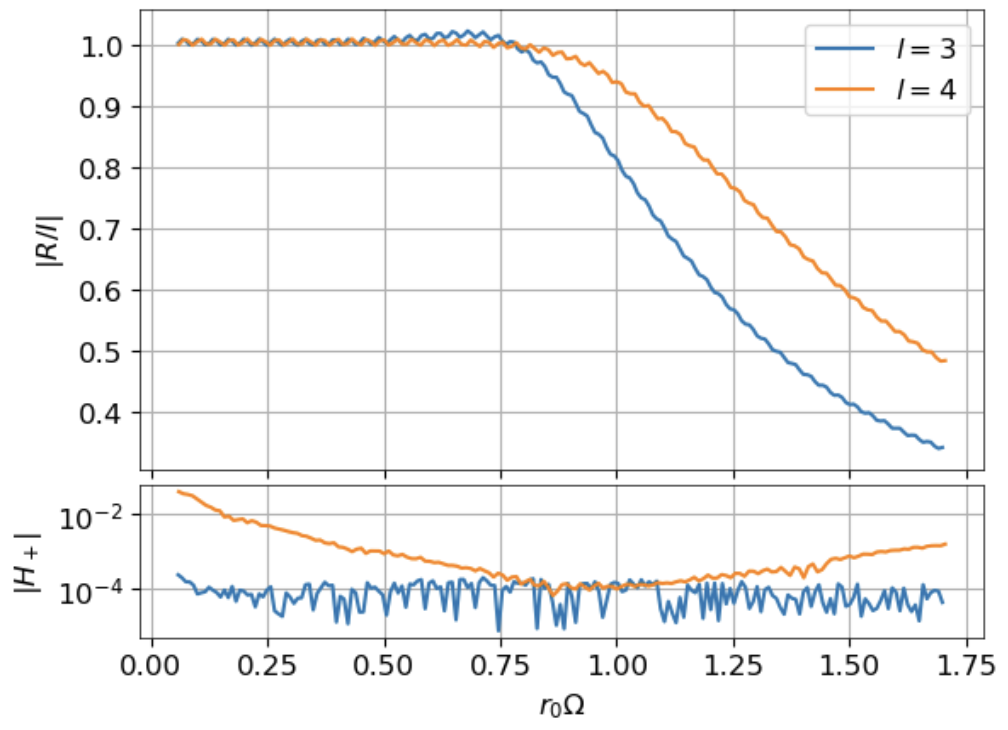}
    \caption{Results for the coefficient $|R/I|$ in the cases $l=3$ and $l=4$. The wiggles in the curves seem to be due to numerical precision issues. The inferior box displays the residual value of the equation.}
    \label{fig:l_34}
\end{figure}

Finally, in figure \ref{fig:l_34} we also show an equivalent computation for higher modes with $l=3,4$. In these cases, we observe how the superradiant effect is suppressed as we increase the value of the angular momentum eigenvalue, most probably due to an increase of the centrifugal barrier. Again, our results are in tune with the conclusions of \cite{Oshita:2021onq}.

The fact that the largest contribution, by far, to superradiant scattering is coming from the $l=1$ mode seems to correspond to an excitation of the vector mode within the gravitational degrees of freedom of the theory \cite{Jacobson:2004ts}. Although here we are not considering a fluctuating space-time, rather working with a fixed background, in a complete computation where backreaction is included, we expect such vector mode to be highly excited by the presence of the scalar field. Importantly, one must note that there is not a linear family of black solutions for a given mass $M$ and different aether configurations, as it might happen in cases where a black hole develops hair \cite{Weinberg:2001gc}. Here, there is no extra quantum number associated to the presence of the aether and, instead, a single regular configuration -- apart from the center singularity -- exists for every $r_0$. Thus, an excitation of the vector mode, such as the one we conjecture here, seems to indicate a strong instability of the configuration and of the universal horizon, pointing out that this space-time is not stable under the evolution of fields with modified dispersion relations, since the black hole cannot decay to a new state. Indeed, this conclusion is in tune with the results of \cite{Rubio:2023eva}.
 
Let us note however that the values of $\epsilon = {\cal N} r_0^2$ -- with ${\cal N}$ a numerical factor -- for which the superradiant effect is large correspond to $\Omega/\Lambda \sim \sqrt{{\cal N}}$ -- since superradiance is found only when $\Omega r_0 \sim {\cal O}(1)$ -- and hence to small energies compared with the Lorentz violating scale, unless $\epsilon \gtrsim 1$. Therefore, as long as the scale $\Lambda$ is at least one order of magnitude larger than the typical energy scale of the black hole $r_{0}^{-1}$, astrophysical size objects described by metric \eqref{eq:space-time} should be meta-stable, since the instability developing time would be large. This is due to the fact that rays returning from the gravitational well back to large radii take a long time to depart form the neighborhood of the Killing horizon \cite{Cropp:2013sea} whenever the ratio $\Omega/\Lambda$ is small.

As a final comment, let us abstract ourselves from the specifics of the construction at hand and highlight that the origin of superradiance within this setting can be traced back to the modified dispersion relation, which allows for the existence of the trapped mode in \ref{fig:characteristics}, that behaves as the usual negative energy mode in the Kerr solution. Its existence is what allows to deplete energy from the black hole configuration. Although here we have coupled our field to EA gravity, there are other settings where modified dispersion relations can appear, such as other possible modified theories of gravity, higher-derivative theories, and, more interestingly, analogue gravity models. The latter actually display a large resemblance with the setting discussed here \cite{DelPorro:2024tuw}, with a similar behavior of modes in the interior of the acoustic horizons of analogue black holes. Thus, we expect a similar superradiant scattering -- and perhaps also instability -- to appear in that case.

\section{Conclusions}\label{sec:conclusions}

In this work, we have studied the scattering of fields equipped with Lorentz violating operators against a black hole geometry. In order for the whole picture to be consistent, we have incorporated LLI violations in our gravitational background by means of EA gravity, and selected a specific Lorentz violating operator in the scalar field action inspired by the construction of Ho\v rava gravity. We have performed a explicit numerical computation in frequency domain and obtained the ratio $|R/I|$ across several values of the Lorentz violating scale $\Lambda$, which we encode here in an effective coupling $\epsilon \sim \Lambda^{-2}$, and of the Killing energy $\Omega$ of the incoming wave.

Our results display superradiance for waves decomposed in spherical harmonics with $l>0$ and $r_0\Omega \sim {\cal O}(1)$, although the effect is suppressed by the centrifugal barrier for larger values of $l$. In particular, the effect is quite strong for the vector mode $l=1$, which can reach values of $|R/I| \sim {\cal O}(10)$ when the Lorentz violating scale is comparable to the size of the black hole.

Altogether, and although astrophysical size objects anjoy a long life-time, this seems to signal a strong instability of the solution, since black holes of this kind have a single quantum number, their mass $M$. The aether configuration surrounding them does not contribute to the total energy of the system \cite{Eling:2005zq}, which means that superradiant scattered waves must be extracting the energy contained in the mass of the configuration, as no other quantity is available. Whether this depletion leads to another smaller black hole or to a totally different configuration is however a complicated question which cannot be answered without resorting to a proper numerical simulation of the system, including back-reaction. This should be possible, as EA gravity is well-posed \cite{Sarbach:2019yso} and suitable for numerical codes, but technical issues plague the problem.

Finally, let us point out that our results here can actually be extrapolated beyond Lorentz violating gravity. Since the key ingredient for the superradiant effect that we observe is the presence of a modified dispersion relation for the scalar field, a similar conclusion should be attainable in any other model where such dispersion relations are possible. This is the case, in particular, of analogue gravity, where acoustic analogues of black holes display properties which are in close resemblance to those discussed along this work \cite{DelPorro:2024tuw}.

\section*{Acknowledgements}
We are grateful to E. Barausse, S. Liberati, F. del Porro, M. Rubio, and M. Schneider for discussions and comments on the draft. The work of M. H-V. has been supported by the Spanish State Research Agency MCIN/AEI/10.13039/501100011033 and the EU NextGenerationEU/PRTR funds, under grant IJC2020-045126-I; and by the Departament de Recerca i Universitats de la Generalitat de Catalunya, Grant No 2021 SGR 00649. IFAE is partially funded by the CERCA program of the Generalitat de Catalunya.


\appendix
\section{Coefficients of the equations} \label{app:coefficients}
We display here the function coefficients of the equation \eqref{ansatz}
\begin{align}
    \alpha_4(r,l)=&\frac{\epsilon  (r_0-2 r)^4}{16 r^5},\\
    \alpha_3(r,l)=&\frac{r_0 \epsilon  (2 r-r_0)^3}{2 r^6},\\
    \alpha_2(r,l)=&-\frac{\epsilon  (2 r-r_0)^2 \left(4 l^2 r^2+4 l r^2+r_0 (16 r-15 r_0)\right)}{8 r^7}-\frac{\left(2 r^2-3 r r_0+r_0^2\right) (2 r-r_0)}{r^2 (r_0-2 r)^2},\\
   \nonumber \alpha_{1}(r,l)=&\frac{\epsilon  (2 r-r_0) \left(8 l^2 r^2 (r-r_0)+8 l r^2 (r-r_0)+r_0 \left(24 r^2-40 r r_0+15 r_0^2\right)\right)}{4 r^8}\\
    &-\frac{r_0 (2 r-r_0) \left(2 i r^2 \omega+2 r-r_0\right)}{r^3 (r_0-2 r)^2},\\
    \nonumber\alpha_0(r,l)=&\frac{\epsilon  \left(-2 l^2 r^2 \left(10 r^2-20 r r_0+7 r_0^2\right)-2 l r^2 \left(12 r^2-20 r r_0+7 r_0^2\right)\right)}{4 r^9}+\frac{\epsilon l^3\left(2 + l\right)}{r^5}+\frac{l \left(l+1\right)}{r^3}\\ 
    & +\frac{\epsilon r_0 \left(-48 r^3+104 r^2 r_0-70 r r_0^2+15 r_0^3\right)}{4 r^9}-\frac{4 r^5 \omega^2-2 i r^3 r_0 \omega-4 r^2 r_0 + 4rr_0^2}{r^4\left(-2r+r_0\right)^2}.
\end{align}

The coefficients in \eqref{eq:eq_tortoise} are, instead
\begin{align}
    \beta_3(r,l)=&\frac{r_0 \left(r_0-2 r\right)}{r^3}\\
    \beta_2(r,l)=&\frac{-4 \left(l^2+l-1\right) r^2 r_0^2 \epsilon -16 l (l+1) r^4 \epsilon +16 l (l+1) r^3 r_0 \epsilon -8 r^6+8 r^5 r_0-4 r r_0^3 \epsilon +r_0^4 \epsilon }{8 r^6 \epsilon }\\
    \beta_2(r,l)=&\frac{-4 \left(l^2+l-1\right) r^2 r_0^2 \epsilon -16 l (l+1) r^4 \epsilon +16 l (l+1) r^3 r_0 \epsilon -8 r^6+8 r^5 r_0-4 r r_0^3 \epsilon +r_0^4 \epsilon }{8 r^6 \epsilon }\\
  \nonumber  \beta_1(r,l)=&\frac{(2 r-r_0)}{8 r^9 \epsilon}\left[ \left(-24 \left(l^2+l-1\right) r^4 r_0\epsilon +4 \left(3 l^2+3 l-13\right) r^3 r_0^2 \epsilon -2 \left(l^2+l-21\right) r^2 r_0^3 \epsilon -4 i r^7 r_0 \omega \right. \right.\\ 
    &+ \left. \left.-2 r^5 \left(r_0^2-8 l (l+1) \epsilon \right) -15 r r_0^4 \epsilon +2 r_0^5 \epsilon \right)\right]\\
   \nonumber \beta_0(r,l)=&\frac{(r_0-2 r)^2}{64 r^{12} \epsilon }\left[ \left(4 l^4 r^4 \epsilon  (r_0-2 r)^2+8 l^3 r^4 \epsilon  (r_0-2 r)^2+2 l^2 r^2 (r_0-2 r)^2 \left(2 r^4-10 r^2 \epsilon +20 r r_0 \epsilon -7 r_0^2 \epsilon \right)\right.\right.\\
   \nonumber&\left.+ 2 l r^2 (r_0-2 r)^2 \left(2 r^4-12 r^2 \epsilon +20 r r_0 \epsilon -7 r_0^2 \epsilon \right)-16 r^{10} \omega ^2+8 i r^8 r_0 \omega +16 r^7 r_0-16 r^6 r_0^2\right) \\ 
    &\left. \left(4 r^5 \left(r_0^3-48 r_0 \epsilon \right)+608 r^4 r_0^2 \epsilon -744 r^3 r_0^3 \epsilon +444 r^2 r_0^4 \epsilon -130 r r_0^5 \epsilon +15 r_0^6 \epsilon \right) \right]
\end{align}

\bibliography{biblio.bib}{}
\end{document}